\documentclass[reprint,superscriptaddress,amsmath, amssymb,aps,prd,notitlepage,longbibliography,floatfix,nofootinbib,twocolumn]{revtex4-1}

\usepackage{amsmath}
\usepackage{lipsum}
\allowdisplaybreaks[4]
\usepackage{cancel}
\usepackage{extarrows}
\usepackage{tensor}     % manipulate tensors
\usepackage{float}
\usepackage[caption = false]{subfig} % combine multiple figures
\usepackage[final]{graphicx}   % include figures
\usepackage[
colorlinks=true,        % color link
citecolor=blue,         % cite color
linkcolor=blue,         % link color
urlcolor=blue           % url color
]{hyperref}             % create hyperlinks
\usepackage{bm}         % \bm{<text>} Bold math symbols
\usepackage{xcolor}     % textcolor
\usepackage{lipsum}
\usepackage{color}      % \textcolor{declared-color}{text}
\usepackage[utf8]{inputenc} % accented characters for .bib file using XeLaTex
\usepackage[section]{placeins} % figures processing
\usepackage{appendix}
\usepackage{units}
\usepackage[capitalise]{cleveref}
\usepackage{units}
\usepackage{orcidlink}
\usepackage[final]{changes}

\newcommand{\nc}{\newcommand*} 
\newcommand{\apjl}{Ap.J.L.}

\nc{\xbar}{\bar{x}}
\nc{\rhoeq}{\rho_{\mathrm{eq}}}
\nc{\zeq}{z_{\mathrm{eq}}}
\nc{\tla}{\tilde{\lambda}}
\nc{\bt}{\beta}
\nc{\dt}{\delta}
\nc{\Dt}{\Delta}
\nc{\vj}{\vec{j}}
\nc{\vl}{\vec{l}}
\nc{\hx}{\hat{x}}
\nc{\hy}{\hat{y}}
\nc{\bj}{\bm{j}}
\nc{\mJ}{\mathcal{J}}
\nc{\mP}{\mathcal{P}}
\nc{\Msun}{M_\odot}
\nc{\app}{\approx}
\nc{\av}[1]{\langle #1 \rangle}
\nc{\eq}[1]{Eq.~\eqref{#1}}
\nc{\al}{\alpha}
\nc{\Xstar}{X_{\ast}}
\nc{\fpbh}{f_{\mathrm{pbh}}}
\nc{\vth}{\vec{\theta}}
\nc{\vla}{\vec{\lambda}}
\nc{\vd}{\vec{d}}
\nc{\Mmin}{M_{\mathrm{min}}}
\nc{\rmd}{\mathrm{d}}
\nc{\mmin}{{m_{\mathrm{min}}}}
\nc{\mmax}{{m_{\mathrm{max}}}}
\nc{\mR}{\mathcal{R}}
\nc{\tmR}{\tilde{\mathcal{R}}}
\nc{\s}{\sigma}
\nc{\ogw}{\Omega_{\mathrm{GW}}}
\nc{\addref}{[\textcolor{red}{add ref}] }
\nc{\Om}{\Omega}
\nc{\gm}{\gamma}
\nc{\Gm}{\Gamma}
\nc{\gpcyr}{\mathrm{Gpc}^{-3}\,\mathrm{yr}^{-1}}
\nc{\Eq}[1]{Eq.~\eqref{#1}}
\nc{\Fig}[1]{Fig.~\ref{#1}}
\nc{\Table}[1]{Table~\ref{#1}}
\nc{\lvc}{LIGO/Virgo} % LIGO-VIRGO collaboration
\nc{\Sec}[1]{Sec.~\ref{#1}}
\nc{\eg}{\textit{e.g.~}}
\nc{\SNR}{\mathrm{SNR}}
\nc{\be}{\mathbf{\epsilon}}
\nc{\bn}{\mathbf{n}}
\nc{\bd}{\mathbf{d}}
\nc{\ba}{\mathbf{a}}
\nc{\eps}{\epsilon}
\nc{\bnu}{\mathbf{\nu}}
\nc{\mb}{\mathbf}
\nc{\bbt}{\mathbf{t}}
\nc{\bth}{\mathbf{\theta}}
\nc{\bep}{\mathbf{\epsilon}}
\nc{\uni}{\mathrm{U}}
\nc{\logu}{\operatorname{\mathrm{log-U}}}
\nc{\RN}{\mathrm{RN}}
\nc{\BN}{\mathrm{BN}}
\nc{\GN}{\mathrm{GN}}
\nc{\mcN}{\mathcal{N}}
\nc{\GWB}{\mathrm{GW}}
\nc{\yr}{\mathrm{yr}}
\nc{\Am}{\mathcal{A}}
\nc{\Dm}{\mathcal{D}}
\nc{\Hm}{\mathcal{H}}
\nc{\sovast}{Soviet Ast.}
\nc{\mrm}{\mathrm}
\nc{\BE}{B\scriptsize{AYES}\normalsize{E}\scriptsize{PHEM}\normalsize  }

\nc{\Ostgw}{\Omega_{\mathrm{GW}}^{\mathrm{ST}}}
\nc{\Ottgw}{\Omega_{\mathrm{GW}}^{\mathrm{TT}}}
\nc{\Ovlgw}{\Omega_{\mathrm{GW}}^{\mathrm{VL}}}
\nc{\Oslgw}{\Omega_{\mathrm{GW}}^{\mathrm{SL}}}
\nc{\cosxi}{\beta}

\nc{\gmPL}{\gamma_{\mathrm{PL}}}
\nc{\APL}{A_{\mathrm{PL}}}

\def\({\left(}
\def\){\right)}
\def\[{\left[}
\def\]{\right]}

\def\e{\begin{equation}}
\def\q{\end{equation}}
\def\m{\begin{eqnarray}}
\def\n{\end{eqnarray}}
\nc{\red}[1]{\textcolor{red}{#1}}

\begin{document}

\title{Targeted search for an individual SMBHB in NANOGrav 15-year and EPTA DR2 data sets}

%%%%%%%%%%%%%%%%%%%%%%%%%%%%%%%%%%%% author %%%%%%%%%%%%%%%%%%%%%%%%%%%%%%%%%%%%
\author{Li-Wen Tian}
\email{tianliwen@itp.ac.cn}
\affiliation{Institute of Theoretical Physics, Chinese Academy of Sciences,Beijing 100190, China}
\affiliation{School of Physical Sciences, 
    University of Chinese Academy of Sciences, 
    No. 19A Yuquan Road, Beijing 100049, China}
%%%%%%%%%%%%%%%%%%%%%%%%%%%%%%%%%%%%
%%%%%%%%%%%%%%%%%%%%%%%%%%%%%%%%%%%%
\author{Yan-Chen Bi\orcidlink{0000-0002-9346-8715}}
\email{Corresponding author: biyanchen@itp.ac.cn}
\affiliation{Institute of Theoretical Physics, Chinese Academy of Sciences,Beijing 100190, China}
\affiliation{School of Physical Sciences, University of Chinese Academy of Sciences, 
No. 19A Yuquan Road, Beijing 100049, China}
%%%%%%%%%%%%%%%%%%%%%%%%%%%%%%%%%%%%
%%%%%%%%%%%%%%%%%%%%%%%%%%%%%%%%%%%% 
\author{Yu-Mei Wu\orcidlink{0000-0002-9247-5155}}
\email{Corresponding author: wuyumei@yzu.edu.cn} 
\affiliation{Center for Gravitation and Cosmology, College of Physical Science and Technology, Yangzhou University, Yangzhou, 225009, China}
%%%%%%%%%%%%%%%%%%%%%%%%%%%%%%%%%%%%  %%%%%%%%%%%%%%%%%%%%%%%%%%%%%%%%%%%%
\author{Qing-Guo Huang\orcidlink{0000-0003-1584-345X}}
\email{Corresponding author: huangqg@itp.ac.cn}
\affiliation{Institute of Theoretical Physics, Chinese Academy of Sciences,Beijing 100190, China}
\affiliation{School of Physical Sciences, 
    University of Chinese Academy of Sciences, 
    No. 19A Yuquan Road, Beijing 100049, China}
\affiliation{School of Fundamental Physics and Mathematical Sciences, Hangzhou Institute for Advanced Study, UCAS, Hangzhou 310024, China}

%%%%%%%%%%%%%%%%%%%%%%%%%%%%%%%%%%%%%%%%%%%%%%%%
%%%%%%%%%%%%%%%%%%%%%%%%%%%%%%%%%%%%%%%%%%%%%%%%%%%%%%

\begin{abstract}
While pulsar timing array (PTA) collaborations have reported evidence for a stochastic gravitational wave background (GWB), the detection of continuous gravitational waves (GWs) from a confirmed supermassive black hole binary (SMBHB) would provide strong support for the SMBHB origin of GWB. 
In this study, we analyze continuous GWs from the SMBHB candidate 3C 66B, 
modeling the GWB as a common uncorrelated red noise.
Using Bayesian methods, we perform targeted searches across two PTA data sets: Nanohertz Observatory for Gravitational Waves 15 years data set and the European Pulsar Timing Array DR2 full data set. 
We find no evidence of such signal in both data sets and then place upper limits on the amplitude of the signal and the chirp mass of the source. 
Additionally, we evaluate the case of a GWB characterized by Hellings–Downs correlations using a likelihood reweighting method, which consistently reconfirms the conclusion of non-detection.

\end{abstract}
% \flushbottom
% \pacs{}
\maketitle

\section{Introduction}

Pulsar Timing Arrays (PTAs) detect nanohertz (nHz) gravitational waves (GWs) by measuring timing residuals—the differences between observed times of arrival (TOAs) from radio pulsar pulses and predictions from a theoretical timing model \cite{Sazhin1978,Detweiler1979,FosterBacker1990}. Recently, PTA collaborations such as the North American Nanohertz Observatory for Gravitational Waves (NANOGrav, \cite{1310.0758,2306.16213}), the European Pulsar Timing Array (EPTA, \cite{1602.08511,2306.16214}), the Parkes Pulsar Timing Array (PPTA, \cite{Zic:2023gta,Reardon:2023gzh}) and the Chinese Pulsar Timing Array (CPTA, \cite{Xu:2023wog}) have reported evidence for an nHz stochastic gravitational wave background (GWB).
The observed GWB, characterized by the Hellings–Downs (HD) correlation \cite{HellingsDowns1983} across pulsars, is consistent with a cosmic population of inspiraling supermassive black hole binaries (SMBHBs), which lose orbital energy through GW emission \cite{Bi:2023tib}. However, alternative scenarios, such as cosmological scalar-induced GWs from primordial black holes \cite{1910.12239,2103.04739}, could also explain the signal \cite{1811.08826,2306.16219,2308.08546,Wu:2023hsa}. Hence, a direct detection of continuous GWs from an individual SMBHB would provide strong evidence for the SMBHB origin of the GWB. With longer observational baselines and improved instrumental precision, such detections are becoming increasingly feasible \cite{2009.11865}, particularly for massive, nearby SMBHBs \cite{1503.04803,2207.01607}.

Moreover, a confirmed GW detection of individual SMBHB would provide critical insights for both theory and observation.
For example, it could illuminate the longstanding ``final parsec problem''—how bound SMBHBs efficiently merge within cosmological timescales \cite{0212270,1702.06964}. 
In addition, multi-messenger observations of such a system would place stringent constraints on the physical processes that govern the merger dynamics \cite{2105.08087,2110.14661}. Finally, an individual SMBHB would serve as a standard siren, enabling independent measurements of cosmological parameters \cite{1903.07644}.
% \replaced{In addition,}{Then} multi-messenger observations of a confirmed SMBHB system could provide stringent constraints on \added{ the physical processes governing} the merger dynamics \cite{2105.08087,2110.14661}.
% And in cosmological studies, a confirmed individual SMBHB could serve as a standard siren\added{, enabling independent measurements of cosmological parameters} \cite{1903.07644}.

While electromagnetic observations could provide evidence of \added{the} existence and properties of SMBHBs candidates in active galaxies \cite{2110.14661, 2111.06882}, typical all-sky searches for individual SMBHB GW signals are \deleted{usually} conducted independently \cite{2306.16222}. However, by incorporating prior knowledge from electromagnetic measurements, so-called targeted searches could enhance the detection probabilities of SMBHBs and \replaced{place tighter upper limits on their parameters}{more precisely constrain their parameter upper limits} \cite{2005.07123, 2105.08087}. These improvements are particularly significant when the SMBHB candidate is located in \added{sky} regions covered by \replaced{a higher density of well-timed pulsars}{more observed pulsars} \cite{2105.08087}.

The radio galaxy 3C 66B with redshift $z=0.020920$ has been proposed to host an SMBHB candidate based on its radio core and jet properties \cite{1502.00632,0106029,0407498,0306103}. 
% Previous electromagnetic and pulsar-timing analyses have produced chirp-mass upper limits spanning $\sim7\times10^8$--$1.3\times10^{10},M_\odot$, with NANOGrav NG11 and NG12.5 yielding limits at the $\sim10^9,M_\odot$ level under various orbital assumptions \cite{0310276,0306103,1011.2647,2005.07123,2309.17438}.
% 3C 66B, an FRI-type radio galaxy with redshift $z=0.020920$ \cite{1502.00632}, resides within the cluster Abell 347, which belongs to the Perseus-Pisces supercluster \cite{0106029,0407498}.
% Some researches have provided evidence for the presence of a SMBHB within this galaxy from its radio core and jet properties \cite{0306103}.
Previous electromagnetic observations placed upper limits on the system's chirp mass of $ 1.3 \times 10^{10} M_\odot$ \cite{0306103} and $7.0 \times 10^8 M_\odot$ \cite{1011.2647} respectively. 
In 2003, the research \cite{0310276} utilizing pulsar timing data from a single pulsar, B1855+09, determines a chirp mass upper limit of $7 \times 10^{9} M_\odot$ under a circular orbit assumption, and noted that a higher upper limit may be possible for an eccentric orbit. 
The NANOGrav Collaboration has specifically investigated potential GW emission from 3C 66B through targeted searches in their data set \cite{2005.07123,2309.17438}.
For a circular orbit, the upper limit on the chirp mass derived from the NANOGrav 11-year data set is $(1.65 \pm 0.02) \times 10^9 M_\odot$ \cite{2005.07123}.
In the NANOGrav 12.5-year data set (hereafter NG12.5), assuming an eccentric orbit, the upper limits are $(1.98 \pm 0.05) \times 10^9 M_\odot$ for the Earth-term-only search and $(1.89 \pm 0.01) \times 10^9 M_\odot$ for the combined Earth-term and Pulsar-term search \cite{2309.17438}.

% In this paper, we follow the pipeline used in the NG12.5 analysis of 3C 66B \cite{2309.17438} and continue the targeted search for the possible SMBHB signal from the source using the NANOGrav 15 years data sets (henceforth NG15) and EPTA DR2 full data sets (henceforth DR2full).
% With modeling the GWB as a common uncorrelated red noise, We find no evidence for such a signal from both data sets. We present the corresponding upper limits on the signal amplitude and the chirp mass of the source.
% Then we compare our result with previous NANOGrav works and extend the discussion to the HD-correlated GWB.
% In Section \ref{sec:source}, we briefly introduce the signal of SMBHBs.  Section \ref{sec:analysis} describes the modelling of NG15 and DR2full data sets as well as Bayesian analysis methods we use. Section \ref{sec:result} presents our main results. And in Section \ref{sec:discussion} we will discuss our results, especially their relation to the HD-correlated GWB.
% We set the $G=c=1$ in this paper.

In this work, we extend the targeted search for a putative SMBHB in 3C 66B using the NANOGrav 15-year (hereafter NG15) and EPTA DR2 full data sets (hereafter DR2full). For computational efficiency, the GWB is modeled as a common uncorrelated red noise. We find no evidence of such a signal in either data set and derive upper limits on the signal amplitude and the chirp mass of the source. We further compare our results with previous NANOGrav studies and examine the impact of an HD-correlated GWB on the analysis.

The structure of the paper is as follows. Section \ref{sec:source} provides a brief overview of SMBHB signals. Section \ref{sec:analysis} describes the modeling of the NG15 and DR2full timing data and the Bayesian analysis methods employed. Section \ref{sec:result} presents the main results, and Section \ref{sec:discussion} discusses their broader implications, particularly when the HD-correlated GWB is considered. Throughout this paper, we adopt units with $G=c=1$.

\section{Source Model}
\label{sec:source}

For pulses propagating from a pulsar $p$ to the Earth, the GW contribution to the residual $R(t)$ is given by
\begin{equation}
R(t) = \int_{t_{0}}^{t} \left( h(t') - h(t' - \Delta_{p}) \right) \rmd t',
\label{eq:res_pri}
\end{equation}
where $h(t)$ denotes the GW strain and $\Delta_p = {d_p}(1 - \cos\mu)$ is the geometric time delay determined by the pulsar distance $d_p$ and the angular separation $\mu$ between the pulsar's line of sight and the GW source direction. The coordinate time $t$ is measured at the solar system barycenter (SSB) according to the DE440 solar system ephemeris, with a fiducial reference time $t_0$ specified at the end of this section. The explicit form of $h(t)$ is
\begin{equation}
    h(t) = 
    \left[ \begin{array}{cc}
    F_{+} & F_{\times}
    \end{array} \right]
    \left[ \begin{array}{cc}
    \cos{2 \psi} & - \sin{2 \psi} \\
    \sin{2 \psi} & \cos{2 \psi} 
    \end{array} \right]
    \left[ \begin{array}{c}
    h_{+}(t) \\
    h_{\times}(t)
    \end{array} \right],
\label{eq:strain_pri}
\end{equation}
where $h_{+,\times}(t)$ are the two GW polarizations in the transverse-traceless gauge, $F_{+,\times}$ are the corresponding antenna pattern functions (see \cite{2210.11454} for their definitions), and $\psi$ is the GW polarization angle.

For convenience, we define
$s_{+,\times}(t) \equiv \int_{t_0}^t h_{+,\times}(t') \rmd t'$,
which allows \Eq{eq:res_pri} to be rewritten in terms of $s_{+,\times}(t)$ as
\begin{equation}
\begin{split}
R(t) &= 
\left[ \begin{array}{cc}
    F_{+} & F_{\times}
    \end{array} \right]
    \left[ \begin{array}{cc}
    \cos{2 \psi} & - \sin{2 \psi} \\
    \sin{2 \psi} & \cos{2 \psi} 
    \end{array} \right]\\
    & \times
    \left[ \begin{array}{c}
    s_{+}(t)-s_{+}(t-\Delta_{p}) \\
    s_{\times}(t)-s_{\times}(t-\Delta_{p})
    \end{array} \right],
\end{split}
\label{eq:res_in_s}
\end{equation}
where $s_{+,\times}(t)$ and $s_{+,\times}(t - \Delta_{p})$ represent the Earth term and Pulsar term, respectively, describing the GW-induced perturbation at the time the pulse is received at Earth and at the time it was emitted from the pulsar.

The explicit form of $s_{+,\times}(t)$ depends on the properties of the GW source.
For an SMBHB system, a general Keplerian-type eccentric orbit is a plausible configuration.
This is also supported by certain theoretical frameworks addressing the final parsec problem, such as the von Zeipel–Lidov–Kozai mechanism \cite{1911.03984,2009.08468,2308.10884,2402.16948,2402.16948,0203370}, which maintains a non-negligible orbital eccentricity until gravitational radiation effectively circularizes the binary.
In this work, we adopt the generalized quasi-Keplerian method \cite{0407049} while neglecting environmental effects.
Our analysis assumes a non-spinning, eccentric, inspiraling SMBHB with slow periastron precession and gradual orbital decay \cite{2309.17438}.
The generalized quasi-Keplerian formalism preserves a Newtonian-like form of the binary’s equations of motion, while incorporating post-Newtonian (PN) corrections \cite{DamourDeruelle1985,SchaferWex1993}.

The quadrupole-order GW signal from an eccentric SMBHB can be derived from the system’s kinematic quantities in the center-of-mass frame, for component masses $m_1$ and $m_2$ \cite{0404128}:
\begin{equation}
\begin{aligned}
    h_+(t)& = -\frac{m \eta}{d_L} \biggl[(1+{c_{\iota}}^2) \left(\frac{m}{r}+r^2 {\dot{\phi}}^2-{\dot{r}}^2\right) \cos 2\phi \\
&+ (1+{c_{\iota}}^2) 2r \dot{r} \dot{\phi} \sin 2\phi \\
&+ {s_{\iota}}^2 \left(\frac{m}{r}-r^2 {\dot{\phi}}^2-{\dot{r}}^2\right) \biggr],\\
    h_{\times}(t)&=-\frac{m \eta}{d_L}\biggl[(2 {c_{\iota}}) (\frac{m}{r}+r^2 {\dot{\phi}}^2-{\dot{r}}^2) \sin 2\phi \\
    &-(2 {c_{\iota}}) 2r \dot{r} \dot{\phi} \cos 2\phi \biggr],
\end{aligned}
\label{eq:strain_dyn}
\end{equation}
where $m=m_1+m_2$ is the total mass, $\eta=m_1 m_2/m^2$ is the symmetric mass ratio, $d_L$ is the luminosity distance, $\iota$ is the orbital inclination with $c_{\iota}=\cos \iota$ and $s_{\iota}=\sin \iota$, $r$ is the binary separation, and $\phi$ is the polar angular coordinate in the orbital plane \cite{9602024}.

For a Keplerian eccentric binary, the orbital motion satisfies
\begin{equation}
\begin{aligned}
    r&=a(1-e \cos u),\\
    l&=nt=u-e \sin u ,\\
    \phi& \equiv f+\omega,\\
    \tan \frac{f}{2}&=\sqrt{\frac{1+e}{1-e}} \tan \frac{u}{2},
\end{aligned}
\label{eq:kepler_newton}
\end{equation}
where $a=\sqrt[3]{m/n^2}$ is the semi-major axis, $n=2 \pi /P$ is the mean motion with orbital period $P$, $e$ is the eccentricity, $l$ is the mean anomaly, $u$ is the eccentric anomaly, $f$ is the true anomaly, and $\omega$ is the argument of periastron.

Using the kinematic relations in \Eq{eq:kepler_newton} and replacing $h_{+,\times}(t)$ with $s_{+,\times}(t)$ by definition, \Eq{eq:strain_dyn} becomes \cite{0110100,2309.17438}
\begin{equation}   
\begin{aligned}
s_{+}(t)=&S_0 \varsigma(t)\big[\left(c_{\iota}^{2}+1\right)\left(-\mathcal{P}\sin2\omega+\mathcal{Q}\cos2\omega\right)\\
& \qquad +s_{\iota}^{2}\mathcal{R}\big],\\
s_{\times}(t)=&S_0 \varsigma(t)(2c_{\iota})\left(\mathcal{P}\cos2\omega+\mathcal{Q}\sin2\omega\right)\,,
\end{aligned}
\label{eq:spx}
\end{equation} 
where the amplitude of $s_{+,\times}(t)$ is separated into $S_0$, its value at $t_0$, and $\varsigma(t)$, describing the subsequent evolution. Here, the subscript “0” denotes $A_0 \equiv A(t_0)$ for a given parameter $A(t)$.
The orbital phase functions $\mathcal{P}$, $\mathcal{Q}$, and $\mathcal{R}$ depend on the eccentricity $e(t)$ and the eccentric anomaly $u(t)$ \cite{2309.17438}:
\begin{equation}   
\begin{aligned}
\mathcal{P}&=\frac{\sqrt{1-e^2} (\cos 2u - e \cos u)}{1-e \cos u},\\
\mathcal{Q}&=\frac{((e^2 -2)\cos u + e) \sin u}{1-e \cos u},\\
\mathcal{R}&=e \sin u.
\end{aligned}
\label{eq:pqr}
\end{equation} 

Considering higher-order effects up to 3PN order, the equations of motion based on \Eq{eq:kepler_newton} are extended to a generalized quasi-Keplerian form \cite{0407049,1707.02088,2210.11454}:
\begin{equation}
\begin{aligned}
    r&=a_r(1-e_r \cos u), \\
    l& \equiv \int_{t_0}^{t} \rmd t' n(t')=u-e_t \sin u+{\mathfrak{F}}_t(u), \\
    \phi&=(1+k)l+{\mathfrak{F}}_{\phi}(u)\\
    &=(1+k)(f-l)+l+\varpi+{\mathfrak{F}}_{\phi}(u), \\
    \tan \frac{f}{2}&=\sqrt{\frac{1+e_{\phi}}{1-e_{\phi}}} \tan \frac{u}{2},\\
    \varpi& \equiv \int_{t_0}^{t} \rmd t' k(t') n(t'),
\end{aligned}
\label{eq:kepler_modified}
\end{equation}
% where the original semi-major axis $a$ and eccentricity $e$ in \Eq{eq:kepler_newton} are replaced by $a_r$ and the set $\{e_t, e_r, e_{\phi}\}$ 
% to achieve the generalized quasi-Keplerian form.
% The expressions for $l$ and $\phi$ include additional terms ${\mathfrak{F}}_t(u)$ and ${\mathfrak{F}}_{\phi}(u)$, respectively; their explicit forms can be found in \cite{0407049, 1707.02088, 2309.17438}.
% Here, the time eccentricity $e_t(t)$ in the quasi-Keplerian formalism is equivalent to the Keplerian orbital eccentricity $e(t)$, and for brevity, we will use $e(t)$ to denote $e_t(t)$ in the following.
% \textcolor{blue}{
% \replaced{
% $k(t)$ is the per-orbit advance of pericenter, which can be expressed in terms of the parameters above; see \cite{1707.02088} for details.
% Finally, $\varpi(t)$ introduced here represents the periastron angle.
% }{
% The quantity \textcolor{red}{$\varpi(t)$ represents the periastron angle, which differs from the argument of periastron $\omega(t)$}.
% Finally, $k(t)$ is the per-orbit advance of pericenter, which can be expressed in terms of the parameters above; see \cite{1707.02088} for details.
% }}
Here, the original semi-major axis $a$ and eccentricity $e$ in \Eq{eq:kepler_newton} are replaced by $a_r$ and the set $\{e_t, e_r, e_{\phi}\}$ to obtain the generalized quasi-Keplerian representation. In this framework, the time eccentricity $e_t(t)$ is equivalent to the Keplerian eccentricity $e(t)$; for convenience, we will simply denote $e_t(t)$ as $e(t)$ hereafter. The mean anomaly $l$ and orbital phase $\phi$ include additional correction terms, ${\mathfrak{F}}_t(u)$ and ${\mathfrak{F}}_{\phi}(u)$. The parameter $k(t)$ represents the per-orbit advance of the pericenter, while $\varpi(t)$ denotes the periastron angle. The explicit forms of all additional parameters and functions introduced in this generalized framework can be found in \cite{0407049,1707.02088,2309.17438}.

\Eq{eq:kepler_modified} describes the conservative quasi-Keplerian dynamics of the binary on short orbital timescales. Over longer timescales, however, the orbital parameters evolve due to gravitational radiation.
Incorporating 2.5PN GW emission, the four parameters $n(t)$, $e(t)$, $\varpi(t)$, and $l(t)$ evolve according to semi-analytic differential equations derived from the general Lagrange method of variation of arbitrary constants \cite{0404128}, given by \cite{2002.03285}:
\begin{equation}
\begin{aligned}
\frac{dn}{dt} & =\frac{1}{5}\left(Mn\right)^{5/3}\eta n^{2}\frac{\left(96+292e^{2}+37e^{4}\right)}{\left(1-e^{2}\right)^{7/2}},\\
\frac{de}{dt} & =\frac{-1}{15}\left(Mn\right)^{5/3}\eta ne\frac{\left(304+121e^{2}\right)}{\left(1-e^{2}\right)^{5/2}},\\
\frac{d\varpi}{dt} & =kn,\\
\frac{dl}{dt} & =n.
\end{aligned}
\label{eq:evolution}
\end{equation}
Given suitable initial values at $t_0$, solving the four equations in \Eq{eq:evolution} uniquely determines the evolution of the eccentric anomaly $u(t)$, which can be computed approximately using Mikkola’s method as a function of $l(t)$ and $e(t)$ \cite{Mikkola1987}.
The argument of periastron $\omega(t)$ is obtained from \Eq{eq:kepler_modified} and the definition $\phi \equiv f+\omega$. Thus, all orbital-phase variables required in \Eq{eq:spx} are fully specified.

We now turn to $S_0$, the amplitude component of $s_{+,\times}(t)$ at $t_0$, its expression including post-Newtonian modifications is \cite{0404128}
\begin{equation}
S_0 = \frac{m\eta}{n_0 d_{\rm L}}\left[m(1+k(m,\eta,n_0,e_0))n_0\right]^{2/3},
\label{eq:S0}
\end{equation}
where $k(m,\eta,n_0,e_0)$ corresponds to $k_0$ for the given values of $m$ and $\eta$.
% The targeted search requires fixed constants parameters, which we align with NG12.5 analysis for consistency \cite{2309.17438}. 
% These parameters, derived separately through electromagnetic observations, include: 
% right ascension, declination and luminosity distance $d_L$ of 3C 66B, the orbital period $P_0$ measured at the fiducial time $t_0$ (set to the observation epoch of \cite{0306103}) to define the initial mean motion $n_0=\frac{2 \pi}{P_0}$ in \Eq{eq:evolution}.

In the targeted search, certain parameters are fixed based on electromagnetic observations.
These include the right ascension, declination, and luminosity distance \textcolor{blue}{\deleted{$d_L$}} of 3C 66B, as well as the orbital period $P_0$ measured at the fiducial time $t_0$ (chosen to coincide with the observation epoch of \cite{0306103}).
The orbital period is then used to define the initial mean motion $n_0 = 2 \pi / P_0$ in \Eq{eq:evolution}.

% To compute the Earth term contributing to residuals, parameters requiring prior distributions specification are:
% the  initial parameters of \Eq{eq:evolution} at $t_0$ (including mean anomaly $l_0$, periastron angle ${\varpi}_0$ and time eccentricity $e_0$), the PTA signal log-amplitude $\log_{10} S_0$ , the symmetric mass ratio $\eta$, the orbital inclination $\iota$, the GW strains polarization angle $\psi$.
% The Pulsar term encodes GW effects at each pulsar's spacetime position during signal reception, introducing additional source orbital phase parameters. 
% Specifically, these parameters—the mean anomaly $l_p$ and periastron angle ${\varpi}_p$ at $t_0$—are treated as independent variables due to current limitations in pulsar distance measurement precision.
% Consequently, they require appropriate prior distributions in the analysis.
% We set these prior distributions of source parameters as the same as NG12.5 analysis \cite{2309.17438}, see in \Table{tb:prior_cw}.
To compute the GW-induced residuals, we specify prior distributions for the initial values in \Eq{eq:evolution} at $t_0$—including the mean anomaly $l_0$, periastron angle ${\varpi}_0$, and time eccentricity $e_0$—as well as for the PTA signal log-amplitude ${\log}_{10} S_0$, the symmetric mass ratio $\eta$, the orbital inclination $\iota$, and the GW strain polarization angle $\psi$.
These parameters are relevant to both the Earth term and the Pulsar term. For the Pulsar term, additional orbital phase parameters are required: the mean anomaly $l_p$ and periastron angle ${\varpi}_p$ at $t_0$.
Due to the limited precision of current pulsar distance measurements, $l_p$ and ${\varpi}_p$ are treated as independent variables.
The pulsar distance $d_p$ is also included as a parameter, with its prior distribution determined following the NG12.5 methodology \cite{2309.17438}: parallax measurements are used when available, and dispersion measure (DM)–based estimates are adopted otherwise.
For DM-based estimates, the conversion to $d_p$ is performed using the \texttt{NE2001p} Python package \cite{2401.05475}, which implements the NE2001 Galactic DM model \cite{0207156}.

\begin{table}[htbp]
\centering
\begin{tabular}{c|c}
\hline
Parameter & \multicolumn{1}{c}{Prior} \\ \hline
 $l_0$&$\uni[0,2\pi]$   \\
 ${\varpi}_0$&$\uni[0,\pi]$ \\
 $e_0$&$\uni[0.001,0.8]\times V$    \\
 $\log_{10} S_0$&$\uni[-12,-6]\times V$  \\
 $\eta$&$\uni[0.001,0.25]\times V$  \\
 $\cos \iota$&$\uni[-1,1]$  \\
 $\psi$&$\uni[0,\pi]$   \\
 $l_p$&$\uni[0,2\pi]$   \\
 ${\varpi}_p$&$\uni[0,\pi]$ \\ \hline
\end{tabular}
\caption{Prior distributions of the source parameters. Here, $V \equiv V(\log_{10} S_0, \eta, e_0)$ denotes a joint probability distribution over $S_0$, $\eta$, and $e_0$, which ensures the validity of the quasi-Keplerian eccentric orbital inspiral model.}
% \caption{\replaced{Prior}{\textcolor{blue}{prior}} distributions of the source parameters, while $V \equiv V(\log_{10} S_0,\eta,e_0)$. The description of prior distribution of $d_p$ is in Section \ref{sec:source} and not \replaced{illustrate}{show} here}
\label{tb:prior_cw}
\end{table}

% The prior for the pulsar distance $d_p$ is set according to the latest measurements, following the same methodology as in NG12.5 analysis \cite{2309.17438}.
% As detailed in \cite{2301.03608}, the distance distribution model prioritizes parallax measurements when available, falling back to DM-based estimates otherwise.
% When employing DM measurements, the conversion to $d_p$ is implemented through the python package \texttt{NE2001p} \cite{2401.05475} using the NE2001 galactic DM model framework \cite{0207156}.

% The inclusion of the Pulsar term parameters $l_p$, ${\varpi}_p$ and $d_p$ reduces computational efficiency.
% The research \cite{2304.03786} based on continues GWs of circular orbital SMBHB concludes that only Earth term analysis could hold a similar constraints on some parameters of SMBHBs with a more efficient calculation.
% So in this paper, we adapt two type of continuous GW signal $R$ in our analysis: one contains both Earth term and Pulsar term (henceforth EPterm) and one only contains Earth term (henceforth Eterm) \cite{2309.17438}.
% We retain the joint probability distribution $V \equiv V(\log_{10} S_0,\eta,e_0)$ in NG12.5 analysis as a criterion in \Table{tb:prior_cw} to constrain prior distributions within a valid region where quasi-Keplerian eccentric orbital inspiral model remains applicable, $V = 1$; outside the region, we enforce $V = 0$ \cite{2002.03285,2309.17438}.

Including the Pulsar term parameters $l_p$, ${\varpi}_p$, and $d_p$ can substantially reduce computational efficiency.
Previous studies on continuous GWs from circular-orbit SMBHBs \cite{2304.03786} have shown that analyses using only the Earth term can achieve comparable constraints on certain SMBHB parameters while being significantly more efficient.
Motivated by this, we consider two signal models: one including both the Earth and Pulsar terms (hereafter EPterm) and one including only the Earth term (hereafter Eterm) \cite{2309.17438}.

For consistency, we fix the same parameters and adopt the same prior distributions for all free parameters as in the NG12.5 analysis \cite{2309.17438}, as summarized in \Table{tb:prior_cw}. To ensure the validity of the quasi-Keplerian eccentric orbital inspiral model, we apply the joint probability distribution $V \equiv V(\log_{10} S_0, \eta, e_0)$ from NG12.5 as a selection criterion: $V = 1$ within the valid parameter region and $V = 0$ outside it \cite{2002.03285,2309.17438}.

\section{Data Analysis}
\label{sec:analysis}
% This section presents the residuals model and Bayesian model comparison framework employed in our analysis.
% We employs python software \texttt{enterprise} \cite{enterprise,2306.16223} and \texttt{enterprise\_extension} \cite{enterpriseextension} for pulsar timing array (PTA) residual calculations and \texttt{GWecc.jl} for continuous GW signal modeling from SMBHB \cite{2002.03285,2210.11454}. For posterior sampling, we utilize \texttt{PTMCMCSampler} \cite{ptmcmc}.

This section outlines the PTA residual model and the Bayesian model-comparison framework used in our analysis.
We employ the Python packages \texttt{enterprise} \cite{enterprise,2306.16223} and \texttt{enterprise\_extension} \cite{enterpriseextension} to compute timing residuals, and \texttt{GWecc.jl} to model continuous GW signals from individual SMBHB \cite{2002.03285,2210.11454}.
Posterior sampling is performed using \texttt{PTMCMCSampler} \cite{ptmcmc}.

\subsection{Timing Residual Model}

% The NG15 data set comprises 16.03 years of observations from the Arecibo Observatory, the Green Bank Telescope (GBT), and the Very Large Array (VLA), with timing data for 68 pulsars \cite{2306.16217}. 
% The EPTA DR2full data set spans 24.7 years of observations, comprising timing data from 25 pulsars collected with six European radio telescopes \cite{2306.16224}. 
% Following the methodology of  NG12.5 analysis, we exclude pulsar J0614–3329 in NG15 data set due to its short observational baseline (less than 3 years) \cite{2309.17438}. 
% Then as in previous research \cite{2309.17438,2005.07123}, we exclude the pulsar PSR J1713+0747 for the presence of two non-GW origin chromatic timing events in both data sets \cite{1712.03651,2306.16218}. 
% Above all, the number of pulsars we use are 66 for NG15 and 24 for DR2full data sets, respectively.

The NG15 data set consists of timing data for 68 pulsars with a total observational span of 16.03 years, collected from the Arecibo Observatory, the Green Bank Telescope (GBT), and the Very Large Array (VLA) \cite{2306.16217}.
The EPTA DR2full data set spans 24.7 years and includes timing data for 25 pulsars obtained with six European radio telescopes \cite{2306.16224}.
Following the methodology of the NG12.5 analysis \cite{2309.17438}, we exclude PSRJ0614$-$3329 from the NG15 data set due to its short observational baseline (less than 3 years).
In addition, consistent with previous studies \cite{2309.17438,2005.07123}, we remove PSRJ1713+0747 from both data sets owing to two chromatic timing events of non-GW origin \cite{1712.03651,2306.16218}.
After these selections, the final data sets contain 66 pulsars for NG15 and 24 pulsars for DR2full.

% We employ the same linearized timing signal model as NG12.5 analysis for NG15 data set \cite{2309.17438}
% \begin{equation}
% \delta t=M \epsilon +n_{\text{WN}}+n_{\text{CRN}}+n_{\text{IRN}}+R,
% \label{eq:ng_model}
% \end{equation}
% and modified the noise model to the same as DR2full GWB search for DR2full data set \cite{2306.16214}
% \begin{equation}
% \delta t=M \epsilon +n_{\text{WN}}+n_{\text{CRN}}+n_{\text{IRN}}+n_{\text{DM}}+n_{\text{SV}}+R,
% \label{eq:epta_model}
% \end{equation}
% where $\delta t$ is the timing residuals,  $M$ is design matrix, $\epsilon$ denotes the difference between the true and estimated timing model parameters vector \cite{1302.1903,2306.16218}, $R$ is the individual SMBHB GW signal of \Eq{eq:res_in_s}. 
% The temporally-uncorrelated white noise component $n_{\text{WN}}$ is characterized through the EFAC and EQUAD parameters, extended with ECORR for NG15 data set analysis in \Eq{eq:ng_model}. 
% $n_{\text{CRN}}$ represents the common red noise background.
% Our analysis employs a common uncorrelated red noise (CURN) model, implemented as a Gaussian process with powerlaw spectral density.
% The CURN process is represented via a Fourier basis spanning $\frac{1}{T}$ to $\frac{x}{T}$, using $x=14$ bins for NG15 and $x=24$ bins for DR2full, with linearly spaced frequencies specific to each data set's observing span.
% These bins numbers are agree with their GWB analysis, respectively \cite{2306.16213,2306.16214}. 
% Here, $T$ denotes the total observational time span of each respective data set.

We model the timing residuals $\delta t$ for the NG15 data set following the NG12.5 analysis \cite{2309.17438} as
\begin{equation}
\delta t=M \epsilon +n_{\text{WN}}+n_{\text{CRN}}+n_{\text{IRN}}+R,
\label{eq:ng_model}
\end{equation}
and for the DR2full data set, we adopt the same noise model used in its GWB search \cite{2306.16214}, given by
\begin{equation}
\delta t = M \epsilon + n_{\text{WN}} + n_{\text{CRN}} + n_{\text{IRN}} + n_{\text{DM}} + n_{\text{SV}} + R,
\label{eq:epta_model}
\end{equation}
where $M$ is the design matrix for the timing model, $\epsilon$ is the vector of differences between the true and estimated parameters \cite{1302.1903,2306.16218}, and $R$ denotes the individual SMBHB GW signal described in \Eq{eq:res_in_s}.

The temporally uncorrelated white noise component $n_{\text{WN}}$ is described by the EFAC and EQUAD parameters, with an additional ECORR term included for the NG15 analysis in \Eq{eq:ng_model}.
% The common red noise background $n_{\text{CRN}}$ is modeled as a common uncorrelated red noise (CURN) process, implemented as a Gaussian process with a power-law spectral density characterized by an amplitude $A_{\text{CRN}}$ and a spectral index ${\gamma}_{\text{CRN}}$.
% \textcolor{blue}{For the NG15 analysis, the priors are $A_{\text{CRN}} \in \logu[10^{-20}, 10^{-11}]$ and ${\gamma}_{\text{CRN}} \in \uni[0,7]$, while in DR2full we take $A_{\text{CRN}} \in \logu[10^{-18}, 10^{-10}]$ and ${\gamma}_{\text{CRN}} \in \uni[0,7]$.}
% The $n_{\text{CRN}}$ term is represented in a Fourier basis covering frequencies from $1/T$ to $N_f/T$, where $T$ is the total observational timespan of the data set and $N_f$ is the frequency bin number. We adopt $N_f=14$ bins for NG15 and $N_f=24$ bins for DR2full, consistent with the respective GWB analyses \cite{2306.16213,2306.16214}.
The GWB is a common red noise $n_{\text{CRN}}$ shared among pulsars, exhibiting correlations according to the HD pattern. However, explicitly encoding this correlation is computationally expensive, so we approximate it with a common uncorrelated red noise (CURN) process. The CURN is described by a power-law spectral density, specified by an amplitude $A_{\text{CRN}}$ and a spectral index ${\gamma}_{\text{CRN}}$, over a Fourier basis spanning frequencies $f \in [1/T, N_f/T]$, where $T$ is the total observation time and $N_f$ is the number of frequency bins. For NG15, we adopt priors $A_{\text{CRN}} \in \logu[10^{-20}, 10^{-11}]$ and ${\gamma}_{\text{CRN}} \in \uni[0,7]$ with $N_f=14$; for DR2full, the priors are $A_{\text{CRN}} \in \logu[10^{-18}, 10^{-10}]$ and ${\gamma}_{\text{CRN}} \in \uni[0,7]$ with $N_f=24$, consistent with previous GWB analyses \cite{2306.16213,2306.16214}.

% For NG15 model in \Eq{eq:ng_model}, $n_{\text{IRN}}$ is the observing-frequency independent individual red noise due to the properties of each pulsar. 
% This term uses the same model as $n_{\text{CRN}}$.  We utilize 30 bins frequency bins and keep the parameters prior distribution of the $n_{\text{IRN}}$ as in NG12.5 analysis \cite{2309.17438}: ${\gamma}_{\text{IRN}} \in \uni[0,7]$ $A_{\text{IRN}} \in \logu[10^{-20}, 10^{-11}]$.

% For the DR2full model in \Eq{eq:epta_model}, $n_{\text{IRN}}$ denotes the  individual red noise as in \Eq{eq:ng_model}, $n_{\text{DM}}$ represents the dispersion measure variations noise, and $n_{\text{SV}}$ corresponds to the scattering variations noise. 
% The $n_{\text{IRN}}$ model maintains the same formulation as in NG15 analysis.
% For $n_{\text{DM}}$ and $n_{\text{SV}}$, we adopt identical functional forms to the $n_{\text{CRN}}$ model but with distinct spectral density representations (see explicit expressions in \cite{2306.16214}).
% All three noise components employ their respective optimal number of Fourier bins as determined in the GWB analysis \cite{2306.16214}.
% We parameterize each noise component with a spectral index ${\gamma}_i$ and amplitude $A_i$, assigning priors ${\gamma}_i \in \uni[0,7]$ and $A_i \in \logu[10^{-18}, 10^{-10}]$ for $i$ in $\text{IRN}, \text{DM}, \text{SV}$.

The term $n_{\text{IRN}}$ denotes the individual red noise arising from intrinsic pulsar rotational irregularities and is independent of the observing frequency $\nu$.
In contrast, the dispersion measure variations $n_{\text{DM}}$ and scattering variations $n_{\text{SV}}$ depend on $\nu$ following a power-law scaling $\nu^{\chi}$.
All three noise components are modeled with a power-law spectrum. For the NG15 analysis, $n_{\text{IRN}}$ is modeled using 30 Fourier frequency bins, with parameter priors identical to those adopted in the NG12.5 analysis \cite{2309.17438}: ${\gamma}_{\text{IRN}} \in \uni[0,7]$ and $A_{\text{IRN}} \in \logu[10^{-20}, 10^{-11}]$.
For the DR2full data set, $n_{\text{IRN}}$, $n_{\text{DM}}$, and $n_{\text{SV}}$ are modeled using the optimal number of Fourier bins determined in the corresponding GWB analysis \cite{2306.16214}.
Each noise component $i \in {\text{IRN}, \text{DM}, \text{SV}}$ is parameterized by a spectral index ${\gamma}_i$ and amplitude $A_i$, with priors ${\gamma}_i \in \uni[0,7]$ and $A_i \in \logu[10^{-18}, 10^{-10}]$.

\subsection{Model Comparison}

% For model comparison, we compute the Bayes factor by  Savage-Dickey formula \cite{Dickey1971,1307.2904}. Below we describe the derivation of this formula.
For model comparison, we evaluate the Bayes factor using the Savage–Dickey formula \cite{Dickey1971,1307.2904}.
The derivation of this formula is outlined below.

% Suppose there are two models ${\Hm}_0$, ${\Hm}_1$ and an observation data set $\Dm$.
% Two models share base parameters $\bm{\theta}$, and the prior distributions of them are the same: $p(\bm{\theta} | {\Hm}_0)=p(\bm{\theta} | {\Hm}_1)$, 
% while ${\Hm}_1$ has some extra parameters $\bm{\zeta}$ which are separable from the prior distributions of the base parameters: $p(\bm{\theta},\bm{\zeta} | {\Hm}_1)=p(\bm{\theta} | {\Hm}_1)p(\bm{\zeta} | {\Hm}_1)$. 
% ${\Hm}_0$ is a nested model within the extended model ${\Hm}_1$, i.e. ${\Hm}_0$ is a special case when $\bm{\zeta}$ of ${\Hm}_1$ have certain values $\bm{\zeta}=\bm{\zeta_1}$, so $p(\Dm|\bm{\theta}, {\Hm}_0)=p(\Dm|\bm{\theta},\bm{\zeta_1},{\Hm}_1)$.

Suppose we have two models, ${\Hm}_0$ and ${\Hm}_1$, and an observational data set $\Dm$.
Both models share a set of base parameters $\bm{\theta}$, with identical priors: $p(\bm{\theta} | {\Hm}_0) = p(\bm{\theta} | {\Hm}_1)$.
Furthermore, model ${\Hm}_1$ includes additional parameters $\bm{\zeta}$, which are assumed to be independent of the base parameters in the prior: $p(\bm{\theta}, \bm{\zeta} | {\Hm}_1) = p(\bm{\theta} | {\Hm}_1) p(\bm{\zeta} | {\Hm}_1)$.
As a result, model ${\Hm}_0$ is nested within ${\Hm}_1$, corresponding to the special case where $\bm{\zeta}$ takes specific values $\bm{\zeta} = \bm{\zeta_1}$ in ${\Hm}_1$.
Accordingly, the likelihood satisfies $p(\Dm | \bm{\theta}, {\Hm}_0) = p(\Dm | \bm{\theta}, \bm{\zeta_1}, {\Hm}_1)$.

The Bayes factor is defined as
\begin{equation} 
\begin{aligned}
\mathcal{B}_{10}&=\frac{p(\Dm|{\Hm}_1)}{p(\Dm|{\Hm}_0)}\\
&=\frac{\int \rmd \bm{{\theta}''} \rmd \bm{{\zeta}''} p(\Dm|\bm{{\theta}''}, \bm{{\zeta}''}, {\Hm}_1) p(\bm{{\theta}''} | {\Hm}_1) p(\bm{{\zeta}''} | {\Hm}_1)}{\int \rmd \bm{{\theta}'} p(\Dm|\bm{{\theta}'}, {\Hm}_0) p(\bm{{\theta}'} | {\Hm}_0)},
\end{aligned}
\label{eq:bayes_sd1}
\end{equation} 
where in the second line we use $p(\bm{\theta},\bm{\zeta} | {\Hm}_1)=p(\bm{\theta} | {\Hm}_1)p(\bm{\zeta} | {\Hm}_1)$.
Multiplying both the numerator and denominator by $p(\bm{\zeta_1}|{\Hm}_1)$ and substituting the terms involving ${\Hm}_0$, we obtain
\begin{equation} 
\begin{aligned}
\mathcal{B}_{10}&=\\
&p(\bm{\zeta_1}|{\Hm}_1) \frac{\int \rmd \bm{{\theta}''} \rmd \bm{{\zeta}''} p(\Dm|\bm{{\theta}''}, \bm{{\zeta}''}, {\Hm}_1) p(\bm{{\theta}''} | {\Hm}_1) p(\bm{{\zeta}''} | {\Hm}_1)}{\int \rmd \bm{{\theta}'} p(\Dm|\bm{{\theta}'},\bm{\zeta_1},{\Hm}_1) p(\bm{{\theta}'} | {\Hm}_1)p(\bm{\zeta_1}|{\Hm}_1)},
\end{aligned}
\label{eq:bayes_sd2}
\end{equation} 
The normalized posterior distribution for the parameters ${\bm{\theta}, \bm{\zeta}}$ under model ${\Hm}_1$ is given by
\begin{equation} 
\begin{aligned}
p(\bm{\theta}, \bm{\zeta}|\Dm, {\Hm}_1)&=\\
&\frac{p(\Dm|\bm{{\theta}}, \bm{{\zeta}}, {\Hm}_1) p(\bm{{\theta}} | {\Hm}_1) p(\bm{{\zeta}} | {\Hm}_1)}{\int \rmd \bm{{\theta}'} \rmd \bm{{\zeta}'} p(\Dm|\bm{{\theta}'}, \bm{{\zeta}'}, {\Hm}_1) p(\bm{{\theta}'} | {\Hm}_1) p(\bm{{\zeta}'} | {\Hm}_1)}.
\end{aligned}
\label{eq:posterior_superior}
\end{equation} 
By evaluating \Eq{eq:posterior_superior} at $\bm{\zeta} = \bm{\zeta_1}$, integrating over $\bm{\theta}$, and replacing the corresponding term in \Eq{eq:bayes_sd2}, we  directly obtain the Savage-Dickey result:
\begin{equation} 
\begin{aligned}
\mathcal{B}_{10}=\frac{p(\bm{\zeta_1}|{\Hm}_1)}{\int \rmd \bm{{\theta}'} p(\bm{{\theta}'}, \bm{\zeta_1}|\Dm, {\Hm}_1)}=\frac{p(\bm{\zeta_1}|{\Hm}_1)}{p(\bm{\zeta_1}|\Dm, {\Hm}_1)},
\end{aligned}
\label{eq:bayes_sd3}
\end{equation} 

% In our models, ${\Hm}_1$ represents two residuals models: \Eq{eq:ng_model} for NG15 and \Eq{eq:epta_model} for DR2full, model ${\Hm}_0$ means residuals model without GW signals from 3C 66B, i.e. without term $R$ in ${\Hm}_1$.
% ${\Hm}_0$ is a nested model within ${\Hm}_1$ whose extra parameter is $S_0=0$.
% Hence the Bayes factor in \Eq{eq:bayes_sd3} is computed using prior $p(S_0=0|{\Hm}_1)$ in numerator and posterior $p(S_0=0|\Dm, {\Hm}_1)$ in denominator. 
% We use samples in the lowest amplitude bin of histogram of $S_0$ as an alternative to ideal $S_0=0$. 
% The uncertainty of Bayes factor is estimated by $\sigma=\frac{\mathcal{B}_{10}}{\sqrt{n}}$ where $n$ is the number of samples in the lowest amplitude bin.

In our analysis, the model ${\Hm}_1$ corresponds to the full residual model: \Eq{eq:ng_model} for NG15 or \Eq{eq:epta_model} for DR2full, whereas ${\Hm}_0$ represents the nested model without GW signals from 3C 66B, i.e., without the term $R$ present in ${\Hm}_1$. In this framework, ${\Hm}_0$ is obtained from ${\Hm}_1$ by setting the extra parameter $S_0$ to zero. Accordingly, the Bayes factor in \Eq{eq:bayes_sd3} is computed using the prior $p(S_0=0|{\Hm}_1)$ in the numerator and the posterior $p(S_0=0|\Dm, {\Hm}_1)$ in the denominator.

In practice, we approximate the ideal value $S_0=0$ by using the samples in the lowest-amplitude bin of the $S_0$ histogram. The associated uncertainty of the Bayes factor is then estimated as $\sigma = \mathcal{B}_{10}/\sqrt{n}$, where $n$ denotes the number of samples in this lowest-amplitude bin.

\section{Result}
\label{sec:result}

\begin{figure*}[!t]%UPDATE!
    \centering 
    \includegraphics[width=0.8\textwidth]{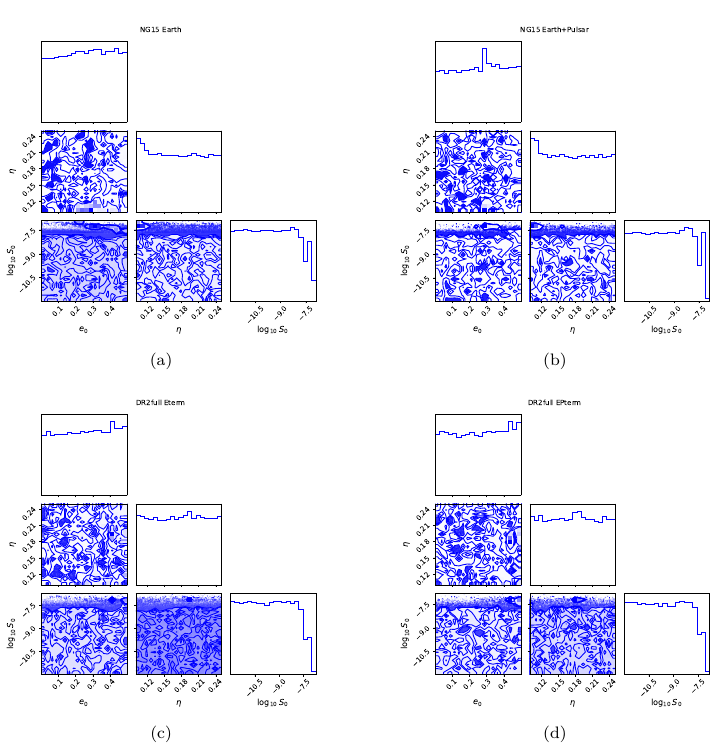}
    % \caption{\textcolor{red}{Posteriors}  $e_0$, $\eta$, and $\log_{10} S_0$ result of NG15 and DR2full search with CURN, (a): NG15 Eterm search; (b): NG15 EPterm search; (c): DR2full Eterm search; (d): DR2full EPterm search.} 
    \caption{Marginalized posteriors of $e_0$, $\eta$, and $\log_{10} S_0$ from the targeted search modeling the GWB as a CURN in the NG15 and DR2full data sets: (a) NG15 Earth term search; (b) NG15 Earth+Pulsar term search; (c) DR2full Earth term search; (d) DR2full Earth+Pulsar term search.}
    \label{fig_curn_posterior} 
\end{figure*}

% We provide no significant evidence for the existence of 3C 66B GW signal in both data sets. 
% For NG15, the Bayes factors are $0.75 \pm 0.01$ for Eterm search and $0.73 \pm 0.02$ for EPterm search. 
% For DR2full, the Bayes factors are $0.78 \pm 0.01$ for Eterm search and $0.78 \pm 0.01$ for EPterm search.
% These results are presented in \Table{tb:BF}.
% For reference, the NANOGrav 11 years \replaced{analysis}{data set} \replaced{reveal}{show} a Bayes factors of $0.74 \pm 0.02$ \cite{2005.07123}.
% NG12.5 \replaced{analysis}{data set} reveal values $0.99 \pm 0.01$ (Eterm) and $1.01 \pm 0.01$ (EPterm) \cite{2309.17438}.
% Our results are consistent with previous research.

We find no significant evidence for a GW signal from 3C 66B in either data set. 
For the NG15 data set, the Bayes factors are measured as 0.81±0.01for both the E-term and EP-term searches. Similarly, for the DR2full data set, the values are 0.78±0.01 for both search types. 
These results are summarized in \Table{tb:BF}. For comparison, the NANOGrav 11-year data set reported a Bayes factor of $0.74 \pm 0.02$ \cite{2005.07123}, while the NG12.5 data set yielded $0.99 \pm 0.01$ (Eterm) and $1.01 \pm 0.01$ (EPterm) \cite{2309.17438}. Taken together, our results are consistent with these earlier studies.

\begin{table}[H]
\centering
% \caption{Bayes factors results for 3C 66B GW signal detection in NG15 and DR2full data sets}
\caption{Bayes factors for the detection of a 3C 66B GW signal in the NG15 and DR2full data sets.}
\begin{tabular}{c|c|c} 
\hline
\multicolumn{1}{c|}{Search} & Eterm & EPterm  \\ 
\hline
NG15     & $0.81 \pm 0.01$ & $0.81 \pm 0.01$   \\ %E:combined; P:combined
DR2full     & $0.78 \pm 0.01$ & $0.78 \pm 0.01$  \\%E:4916; P:4918
\hline
\end{tabular}
\label{tb:BF}
\end{table}

\begin{table*}[!t]
\centering
\begin{tabular}{l|cc|cc}
\hline
 & \multicolumn{2}{c|}{$S_0/\text{ns}$}    & \multicolumn{2}{c}{$M_c/{10}^9 M_\odot$}    \\ \hline
 & \multicolumn{1}{c|}{Eterm} & EPterm & \multicolumn{1}{c|}{Eterm} & EPterm \\ \hline
NG15 & \multicolumn{1}{c|}{$80.88 \pm 2.16$} & \multicolumn{1}{c|}{$75.66 \pm 2.54$} & \multicolumn{1}{c|}{$1.89 \pm 0.03$} & \multicolumn{1}{c}{$1.82 \pm 0.03$} \\ %E:combined; P:combined
DR2full & \multicolumn{1}{c|}{$96.27 \pm 2.77$} & \multicolumn{1}{c|}{$85.37 \pm 2.18$} & \multicolumn{1}{c|}{$2.08 \pm 0.04$} & \multicolumn{1}{c}{$1.95 \pm 0.03$} \\ %E:4916 P:4919
NG12.5 & \multicolumn{1}{c|}{$88.1 \pm 3.7$} & \multicolumn{1}{c|}{$81.74 \pm 0.86$} & \multicolumn{1}{c|}{$1.98 \pm 0.05$} & \multicolumn{1}{c}{$1.89 \pm 0.01$} \\ \hline
\end{tabular}
% \caption{\replaced{95\% upper limits results}{\textcolor{blue}{95\% Upper Limits Results}} for 3C 66B $S_0$ and $M_c$ in NG15 and DR2full data sets. The NG12.5 analysis results in \cite{2309.17438} are also \replaced{illustrated}{shown} here}
\caption{$95\%$ upper limits on the signal amplitude $S_0$ and chirp mass $M_c$ for 3C 66B derived from the NG15 and DR2full data sets. Results from the NG12.5 analysis \cite{2309.17438} are included for comparison.}
\label{tb:post_ul}
\end{table*}

In our analysis, we first obtain quasi-posteriors for the parameters $e_0$, $\eta$, and $S_0$ using the priors listed in \Table{tb:prior_cw}, without applying the validity criterion $V$. These quasi-posteriors are then refined by accounting for the interdependence specified by $V$ to derive robust upper limits on $S_0$ and the chirp mass $M_c = m \eta^{3/5}$. Following the pipeline of \cite{2309.17438}, we introduce a threshold criterion to identify regions of parameter space where the observational data meaningfully constrain the posterior. Specifically, the $(e_0, \eta)$ plane is divided into uniformly spaced pixels. A pixel is considered valid only if the systematic difference between the prior- and posterior-derived $95\%$ upper limits of $\log_{10} S_0$ exceeds $5\%$; otherwise, it is regarded as astrophysically insignificant and discarded. The set of valid pixels then defines the region of parameter space considered reliable for inference.

When compared with the valid region boundary obtained in the NG12.5 analysis, the NG15 results show a tendency toward slightly tighter constraints on $e_0$, accompanied by somewhat looser limits on $\eta$. In contrast, the DR2full analysis exhibits a mild preference for weaker $e_0$ constraints while maintaining $\eta$ bounds similar to those in NG12.5. Nevertheless, the valid region defined by $\eta > 0.1$ and $e_0 < 0.5$ in the NG12.5 analysis \cite{2309.17438} remains applicable across all data sets considered here. We therefore adopt this criterion throughout our analysis.

Within this valid region, we present the marginalized posterior distributions for $e_0$, $\eta$, and $\log_{10} S_0$ from all data sets in \Fig{fig_curn_posterior}. The corresponding $95\%$ upper limits for $S_0$ and $M_c$ are computed exclusively from data within this region, and the results are summarized in \Table{tb:post_ul} together with those from the NG12.5 analysis \cite{2309.17438}. As in \cite{2309.17438}, we evaluate the uncertainties in the upper limits using the bootstrap method.

For NG15, the upper limits are $S_0 = 80.88 \pm 2.16 \ \text{ns}$ and $M_c = (1.89 \pm 0.03) \times 10^9 \ M_\odot$ for the Eterm search, and $S_0 = 75.66 \pm 2.54 \ \text{ns}$ and $M_c = (1.82 \pm 0.03) \times 10^9 \ M_\odot$ for the EPterm search. For DR2full, the corresponding values are $S_0 = 96.27 \pm 2.77 \ \text{ns}$ and $M_c = (2.08 \pm 0.04) \times 10^9 \ M_\odot$ for the Eterm search, and $S_0 = 85.37 \pm 2.18 \ \text{ns}$ and $M_c = (1.95 \pm 0.03) \times 10^9 \ M_\odot$ for the EPterm search.
Overall, the NG15 results remain consistent with those from NG12.5, though they yield marginally tighter constraints on both $S_0$ and $M_c$. By contrast, the DR2full results indicate less stringent upper limits relative to the other data sets.

%We also check the posterior distributions for the CURN parameters.
%The results consistently exhibit nearly flat posterior distributions, indicating that our analysis cannot extract meaningful parameter constraints.
%This behavior is consistent with the findings reported in NG12.5 search \cite{2309.17438}.
%We check the results of CURN parameters, and they are in consistent with the GWB search results which indicate the pipeline is still robust when searching 3C 66B continuous GW signal in new data sets \cite{2309.17438}. 
%They all in agree with the model containing no signal from the 3C 66B, respectively. 

\section{Conclusion and Discussion}
\label{sec:discussion}

% In this paper, we perform a targeted search for potential continuous GW signal from the SMBHB candidate 3C 66B in NG15 and EPTA DR2full data sets \cite{2309.17438,2306.16213, 2306.16214}. 
% We find no such GWs by using Bayesian analysis and then place the 95\% upper limit on chirp mass $M_c$ and signal amplitude at fiducial time $S_0$.
% The Bayes analysis results are presented in \Table{tb:BF}, and the upper limits results are exhibited in \Table{tb:post_ul} with the NG12.5 analysis upper limits results in \cite{2309.17438}.
In this work, we conduct a targeted search for continuous GW signals from the SMBHB candidate 3C 66B using the NG15 and EPTA DR2full data sets \cite{2309.17438,2306.16213,2306.16214}.
Applying Bayesian analysis, we find no evidence for such signals and consequently derive $95\%$ upper limits on the chirp mass $M_c$ and the fiducial signal amplitude $S_0$.
The Bayes factor results are summarized in \Table{tb:BF}, while the derived upper limits are reported in \Table{tb:post_ul}, together with the corresponding constraints from the NG12.5 analysis \cite{2309.17438}.

The above results are based on modeling the GWB as a CURN, whereas both data sets actually favor an HD-correlated GWB over CURN \cite{2306.16213,2306.16214}. Although computational constraints prevent us from directly implementing a full HD-correlated common red noise in the search, we can instead use a likelihood reweighting technique to estimate the evidence for an HD-correlated GWB, as previously applied in the NG15 all-sky search for individual SMBHBs \cite{2306.16222}.

Based on the correlation structure of $n_{\text{CRN}}$ and the presence of $R$ in the residual models \Eq{eq:ng_model} and \Eq{eq:epta_model}, we consider four models: ${\Hm}_{\text{c}}$ (CURN only), including only a common uncorrelated red-noise process; ${\Hm}_{\text{h}}$ (HD only), including only an HD-correlated GWB; ${\Hm}_{\text{cr}}$ (CURN + SMBHB), combining a CURN process with an individual SMBHB signal; and ${\Hm}_{\text{hr}}$ (HD + SMBHB), combining an HD-correlated GWB with an individual SMBHB signal.
The likelihood reweighting method evaluates Bayes factors for ${\Hm}_{\text{hr}}$ versus ${\Hm}_{\text{cr}}$ using posterior samples from ${\Hm}_{\text{cr}}$. For a given set of samples, weights are defined as the ratio of the likelihood under ${\Hm}_{\text{hr}}$ to that under ${\Hm}_{\text{cr}}$, and the Bayes factor is estimated as the average of these weights \cite{2212.06276}. In our analysis, we reweight posterior samples from $\mathcal{H}_{\text{cr}}$ obtained in Section \ref{sec:result}. For NG15, the Bayes factors of ${\Hm}_{\text{hr}}$ versus ${\Hm}_{\text{cr}}$ are 1.00 in both Eterm and EPterm searches; for DR2full, they are 1.51 (Eterm) and 1.50 (EPterm), all with uncertainties below 0.01.
\textcolor{blue}{\deleted{We find that the posterior distributions of the SMBHB parameters in \Fig{fig_curn_posterior} remain largely unchanged when incorporating HD-correlations.}}
In addition, from the NG15 and EPTA DR2full GWB analyses, the Bayes factors of ${\Hm}_{\text{h}}$ versus ${\Hm}_{\text{c}}$ are 226 and 4, respectively \cite{2306.16213,2306.16214}. Combining these with the ${\Hm}_{\text{cr}}$ versus ${\Hm}_{\text{c}}$ results from Section \ref{sec:result}, we further derive the Bayes factors of ${\Hm}_{\text{hr}}$ versus ${\Hm}_{\text{h}}$. For NG15, this ratio is below 0.01 in both Eterm and EPterm searches, while for DR2full it is 0.29 for both. Compared to the previous ${\Hm}_{\text{cr}}$ versus ${\Hm}_{\text{c}}$ results, the Bayes factors for ${\Hm}_{\text{hr}}$ versus ${\Hm}_{\text{h}}$ are substantially suppressed. Similar reductions are observed in the NG15 and EPTA DR2 all-sky analyses. This effect likely arises because the GWB and SMBHB signals are partially covariant, since both contribute to the HD correlation \cite{1305.0326,2306.16226}, as indicated by the fact that the evidence for an HD-correlated GWB is also significantly reduced when an SMBHB signal is included \cite{2306.16222}.
While the NG15 all-sky search reveals tentative evidence for two candidate GW signals at $\sim 4$ nHz and $\sim 170$ nHz \cite{2306.16222}, and the EPTA DR2 analysis also identifies one significant signal at $\sim 4$ nHz \cite{2306.16226}, both results are obtained under the CURN framework, and the evidence diminishes when HD correlations are included in the CRN model. Notably, neither candidate appears near the expected frequency of 3C 66B. Moreover, as the Bayes factors (${\Hm}_{\text{hr}}$ versus ${\Hm}_{\text{h}}$) remain consistently below 1, our conclusion of no detectable GW signal from 3C 66B remains unchanged.

\textit{Acknowledgements.}
This work is supported by the National Key Research and Development Program of China Grant No.2020YFC2201502, the grants from NSFC (Grant No.~12475065, 12447101) and the China Manned Space Program with grant no. CMS-CSST-2025-A01. We acknowledge the use of HPC Cluster of ITP-CAS.

\bibliography{refs}
\end{document}